\documentclass{aa}

\begin{document}    
    
    
\title{Comment on the first-order Fermi acceleration at
          ultra-relativistic shocks}
 
\author{M. Ostrowski\inst{1,}\inst{2} \and J. Bednarz\inst{1}}
 
\institute{Obserwatorium Astronomiczne, Uniwersytet Jagiello\'nski,
                   ul. Orla 171, 30-244 Krak\'ow, Poland \and
Institut f\"ur Theoretische Physik, Lehrstuhl IV: Weltraum und
Astrophysik, Ruhr-Universit\"at Bochum, Germany} 
 
\offprints{M. Ostrowski, mio\@@oa.uj.edu.pl}
    
\date{Received 17 April 2002 / Accepted 23 July 2002}

\titlerunning{The first-order Fermi acceleration at
          ultra-relativistic shocks}

\authorrunning{M. Ostrowski \& J. Bednarz}
    
\abstract{
The first-order Fermi acceleration process at an ultra-relativistic
shock wave is expected to create a particle spectrum with the unique
asymptotic spectral index $\sigma_{\gamma \gg 1} \approx 2.2$~. Below,
we discuss this result and differences in its various derivations, which
-- explicitly or implicitly -- always require highly turbulent
conditions downstream of the shock. In the presence of medium amplitude
turbulence the generated particle spectrum can be much steeper than the
above asymptotic one. We also note problems with application of the
pitch angle diffusion model for particle transport near the
ultra-relativistic shocks.
     }

\maketitle
 
\section{Introduction} 
 
Ultra-relativistic shock waves suggested to be sources of gamma-ray
bursts are also expected by some authors to produce ultra-high-energy
cosmic ray particles. A process of the first-order Fermi acceleration in
such shocks was discussed in a series of papers by Bednarz \& Ostrowski
(1997, 1998; see also Bednarz 2000a,b), Gallant \& Achterberg (1999; see
also Achterberg et al. 2001), Kirk et al. (2000) and Vietri (2002).
Below, in section 2, we briefly compare and discuss different approaches
to the considered acceleration process, leading to the asymptotic
spectral index $\sigma_{\gamma \gg 1} \approx 2.2$. We note an important
fact that in order to derive this result, essentially all these studies
consider the large amplitude magnetic field perturbations near the
shock, with the turbulence power concentrated in the short wavelength
range. The particle energy spectra generated in shocks propagating in a
mildly turbulent medium, with the limited turbulence downstream of the
shock, can be much steeper in ultra-relativistic shocks than the above
`asymptotic' one.

In the discussion below we neglect the strictly parallel shocks, where
some of our objections can be invalid. However, such shocks are not
expected to frequently occur in the universe.

\section{On the first-order Fermi acceleration at ultra-relativistic 
shocks} 
 
The first-order Fermi acceleration process at an ultra-relativistic
shock wave involves extreme particle anisotropy at the shock in the
upstream plasma rest frame (UPF), and more mild distributions in the
shock normal rest frame or the downstream plasma rest frame (cf.
Begelman \& Kirk 1990). Let us consider an individual cosmic ray
particle acceleration starting with a particle crossing the shock
upstream (cf. a detailed discussion by Gallant \& Achterberg 1999).
Then, in UPF, its momentum is nearly parallel to the shock normal. When
the shock Lorentz factor is large ($\gamma \gg 1$) the particle moves in
front of the shock for a time required for a slight deflection of its
momentum allowing the shock to overtake it and transmit to the
downstream region. The deflection proceeds due to the magnetic field
upstream of the shock, consisting of the large scale smooth background
structure perturbed by the MHD fluctuations. This tiny change of
particle momentum upstream of the shock allows for its transmission
downstream of the shock, where -- due to the Lorentz transformation with
a large $\gamma$ -- its momentum direction can be changed at a large
angle with respect to its original direction before the transmission
upstream. Such large amplitude angular scatterings can enable a finite
fraction of particles to follow trajectories leading to the successive
transmissions upstream of the shock. Repeating of the described loops,
with each roughly doubling the particle energy, leads to formation of
the power law particle spectrum. Several authors (Bednarz \& Ostrowski
1998, Gallant \& Achterberg 1999, Gallant et al. 1999) discussed this
process leading to formation of the spectrum with the energy spectral
index $\sigma \approx 2.2$ at $\gamma \gg 1$~. Essentially the same
results were obtained within different approaches presented by the above
authors and by Kirk et al. (2000) and Vietri (2002). We do not consider
here the papers postulating the acceleration process, but not discussing
details of the proposed mechanism.

The work of Bednarz \& Ostrowski (1997, 1998) was based on Monte Carlo
simulations of particle transport governed by small amplitude pitch
angle scattering. Thus, depending on the scattering parameter $\Delta t$
(a mean time between successive scattering acts) and $\Delta
\Omega_{max}$ (the maximum angular scattering amplitude), we were able
to model situations with different mean field configurations and
different amounts of turbulence. One should note that the mean field
configuration downstream of the shock was derived here from the mean
upstream field using the appropriate jump conditions and trajectories of
particles interacting with the shock discontinuity were derived exactly
for such fields. A particle trajectory was derived in the respective
local plasma rest frame, with the Lorentz transformation applied at each
particle shock crossing. The approach takes into account correlations in
the process due to the regular part of the magnetic field, but
irregularities responsible for pitch angle scattering are introduced at
random. In order to model particle pitch angle diffusion upstream of the
shock, with nearly a delta-like angular distribution $\theta \sim
\gamma^{-1}$ ($\theta$ - a momentum vector inclination to the shock
normal), an extremely small scattering amplitude should be used, $\Delta
\Omega_{max} \ll \gamma^{-1}$\footnote{ The condition $\Delta
\Omega_{max} \ll \gamma^{-1}$ lead to the excessive computation
times. Thus, in our simulations we used a relatively `large' maximum
amplitude $\Delta \Omega_{max} = {1 \over 2} \gamma^{-1}$, but
comparison to the results obtained with smaller $\Delta \Omega_{max}$
revealed only insignificant differences in the obtained particle spectra
(as measured at the escape boundary placed at $4 r_g$ downstream of the
shock; $r_g$ is a particle gyroradius). One should note that due to the
above choice of $\Delta \Omega_{max}$ and because in our numerical code
the relative particle velocity with respect to the shock instead of the
velocity in the plasma rest frame was improperly used for the particle
weighting function in the plasma rest frame, the angular distributions
presented by us (Bednarz \& Ostrowski 1998) are slightly different in
comparison to the results of Gallant et al. (1999) and Kirk et al.
(2000). However, this error leads only to a wrongly presented angular
distributions and it does not influence particle distributions
considered in simulations and the derived spectra.}. Increasing the
shock Lorentz factor results in decreasing the momentum perturbation
required for its transmission downstream and leaves a shorter time for
this perturbation, $t_1$. In the applied pitch angle diffusion approach,
the momentum variation due to the regular component of the magnetic
field scales like $t_1$, whence the diffusive change scales like
$t_1^{1/2}$. Thus growing $\gamma$ leads to decreasing $t_1$ and the
diffusive term has to dominate at sufficiently large $\gamma$. It is
the reason why in our simulations the orientation of the regular
magnetic field ceases to play a role in the limit $\gamma \to \infty$,
resulting in the spectral index convergence to its asymptotic value.

However, one should note that with decreasing $\Delta t$ and $\Delta
\Omega_{max}$, when the interaction proceeds at the sub-resonance ($\ll
r_g$) spatial scale, a serious physical problem with the applied
approach appears. In order to scatter particle momentum {\it uniformly}
within a narrow cone centred on the initial momentum, it requires the
short wave turbulence to be non-linear at the shortest scales. In our
discussion of the `effective' magnetic field, $B_e$, in the pitch angle
diffusion simulations (Bednarz \& Ostrowski 1996; cf. Appendix below) we
evaluate the lower limit of such an effective field from the curvature of
simulated particle trajectories as

$$B_e = B_o \sqrt{ 1 + \left( 0.67 {\Delta \Omega_{max} \over \Delta t}
\right)^2} \qquad , \eqno(1)$$

\noindent
taking into account both the background uniform field $B_o$ and the
turbulent component evaluated with the use of the scattering parameters
$\Delta \Omega_{max}$ and  $\Delta t$ (in this expression $\Delta t$ is
given in angular units, it stands for $c \Delta t / r_g(B_o)$). Assuming
the constant pitch angle diffusion coefficient ($\propto (\Delta
\Omega_{max})^2 / \Delta t$) for a series of computations involving
smaller and smaller $\Delta \Omega_{max} \sim \gamma^{-1}$ we had to use
$\Delta t$, which scales like $\gamma^{-2}$. As a result, to be
consistent with the assumed scattering model, for large shock Lorentz
factors the effective magnetic field increases to large values due to
the required growing power being
concentrated in the short wave turbulence, $B_e
\propto \gamma B_o $. Such conditions seem to be unrealistic at least
upstream of the shock.

An analogous pitch angle diffusion modelling appended the considerations of
Gallant et al. (1999; for a more detailed description see Achterberg et
al. 2001). They considered the highly turbulent conditions near the shock
leading to the particle pitch angle diffusion {\it with respect to the
shock normal}, i.e. the regular part of the magnetic field was
neglected. These computations gave essentially the same spectral indices
as the asymptotic one derived by Bednarz \& Ostrowski (1998). Also, in a
variant of this model with uniform magnetic field upstream of the shock
and fully chaotic turbulent field downstream, the resulting spectral
index did not vary substantially. The physical content of the
discussed model is substantially different from the Bednarz \&
Ostrowski one because it neglects the influence of the uniform field (or
long wavelength perturbations with $\lambda > r_g$) resulting in
magnetic field correlations at both sides of the shock. Thus, it
provides spectra with the asymptotic spectral index at quite moderate
$\gamma \sim 10$, a feature also present in the Bednarz \& Ostrowski
simulations for parallel shocks. However, if the amplitude of the
magnetic field turbulence is limited, these simulations cannot
reproduce spectrum steepening (or flattening at intermediate Lorentz
factors) in the presence of oblique magnetic fields (cf. Ostrowski 1993,
Bednarz \& Ostrowski 1998, Begelman \& Kirk 1990). Both the above models
describe essentially the same physical situation for shocks
propagating in the highly turbulent medium and, of course, in rarely --
if ever -- occurring parallel relativistic shocks.

An alternative discussion of the acceleration process presented by
Gallant \& Achterberg (1999) was based on a simple turbulence model. In
their approach a highly turbulent magnetic field configuration was
assumed upstream and downstream of the shock, idealized as cells filled
with randomly oriented, uniform (within a cell) magnetic fields. With
such an approach, particles crossing the shock enter a new cell with a
randomly selected magnetic field configuration. Thus, there always occur
configurations allowing some particles crossing downstream to reach the
shock again and again. As a result of successive energy increases of the
same finite fraction of accelerated particles, the power law spectrum is
formed. In this model there is no need for upstream magnetic field
perturbations if the considered oblique magnetic field configuration can
turn all upstream particles back to the shock.

Two quasi-analytic approaches to the considered acceleration process
were presented by Kirk et al. (2000) and Vietri (2002). Both provide
methods to solve the Fokker-Planck equation describing particle
advection with the general plasma flow and the small amplitude
scattering of particle pitch angle as measured with respect {\it to the
shock normal}. The important work of Kirk et al. modified the Kirk \&
Schneider (1987) series expansion approach to treat the delta-like
angular distribution upstream of the shock. An analytically more simple
Vietri approach applies convenient ansatz'es for the anisotropic
upstream and downstream particle distributions, resembling the Peacock
(1981) approach to acceleration at `ordinary' relativistic shocks. Both
methods confirm the results of the earlier numerical modelling. A
deficiency of the above semi-analytic approaches is its inability to
treat situations with mildly perturbed magnetic fields, on average
oblique to the shock normal. If considered valid for different
configurations of the mean magnetic field, these models require the large
amplitude short wave turbulence to remove signatures in particle
trajectories of the uniform background field or of the long wave
perturbations. Thus it provides an alternative description of the same
physical situation discussed earlier with numerical methods by Bednarz
\& Ostrowski (1998) in the $\gamma \to \infty$ limit or their parallel
shock results and all other authors applying small amplitude pitch angle
scattering simulations at parallel shock waves.

\section{Conclusions}

The discussed approaches to the
cosmic ray first-order Fermi acceleration at relativistic shocks yield
consistent estimates of the asymptotic spectral index $\approx (2.2, \,
2.3)$. However, the result is not as universal as one could infer from
the convergent conclusions of different authors, because all presented
derivations require (explicitly or implicitly) large amplitude MHD
turbulence near the shock. Only the Bednarz \& Ostrowski (1998) modelling
allows one to treat -- in a simplified way -- conditions with medium
amplitude perturbations of the magnetic field. In such conditions
particle spectra are expected to be very steep at high shock Lorentz
factors. The spectra considered by Bednarz \& Ostrowski flatten at large
$\gamma$ due to an implicit increase of the short wave turbulence in
their model, approaching closer and closer the parallel shock
configuration considered by the other authors. Until now the situation
with the medium amplitude turbulence, $\delta B < B_o$, has not been
studied
in the limit of large $\gamma$, however, from comparison with the
results of Begelman \& Kirk (1990), Ostrowski (1993) and of Bednarz \&
Ostrowski (1998) for intermediate shock Lorentz factors, we expect
very steep spectra to be formed in such conditions. Thus, if the
conditions with limited turbulence are met at a large $\gamma$ shock, it
can be unable to accelerated particles to very high energies in the
first-order Fermi mechanism. On the other hand the `low' energy
electrons radiating from ultra-relativistic shocks could be accelerated
by the non-first-order processes, analogous to the ones discussed by
Hoshino et al. (1992) or Pohl et al. (2001).

The main deficiency of the approaches applying the pitch angle diffusion
equation, in particular of our own attempt to discuss cases with oblique
background magnetic fields, is their limitation to particular, highly
turbulent conditions near the shock. This limitation may be significant
or non-significant, depending on whether such conditions exist
downstream of the ultra-relativistic shock. In the process discussed by
Medvedev \& Loeb (1999) such short-wave non-linear turbulence is created
downstream of the shock, in the non-resonant wave-vector range for the
shock accelerated particles.

\begin{acknowledgements}    
 
MO is grateful to Reinhard Schlickeiser for the invitation to the Institute
for Theoretical Physics of the Ruhr University, where this work was
partly done, and to Yves Gallant and Bohdan Hnatyk for valuable
discussions. The work was supported by the {\it Komitet Bada\'n
Naukowych} through the grant PB 258/P03/99/17.
 
\end{acknowledgements} 

\section*{Appendix: Evaluation of the effective magnetic field $B_e$ in
a numerical code applying the discrete small-amplitude pitch angle
scattering method}

Let us evaluate the magnetic field components responsible for regular,
$\Delta \theta_r$, and turbulent, $\Delta \theta_t$, angular deviations
of the particle momentum during a single particle propagation time-step
$\Delta t$. For the regular deviation due to the mean magnetic field
$B_o$

$$\Delta \theta_r \sim {c \Delta t \over r_g(B_o)} = {e \over p} B_o
\Delta t \qquad , \eqno(A1)$$

\noindent
where $r_g(B) = e B / (p c)$, $p$ is the relativistic particle momentum
and $e$ its charge. In the above estimate the undefined proportionality
factor depends on local particle distribution anisotropy.  For the
turbulent component of the magnetic field one can evaluate its {\it
lower limit} analogously to (A1) by assuming the turbulent field
component, $B_t$, to be uniform at the spatial scale $c \, \Delta t$:

$$\Delta \theta_t \sim {c \Delta t \over r_g(B_t)} = {e \over p} B_t
\Delta t \qquad . \eqno(A2)$$

\noindent
For the considered model involving uniform scattering within a narrow
cone of the opening angle $\Delta \Omega_{max} \ll 1$, the mean
scattering angle equals $\frac{2}{3} \Delta \Omega_{max}$. If the
magnetic field components $\vec{B}_o$ and $\vec{B}_t$ are oriented
randomly with respect to each other, then the effective field modifying
particle trajectory, $B_e$, can be evaluated as $B_e^2 = B_o^2 + B_t^2$.
With the estimate ${e \over p} B_t\Delta t  = \frac{2}{3} \Delta
\Omega_{max}$ one obtains $B_t = B_o \{ \frac{2}{3} \Delta \Omega_{max} /
[(e B_o / p) \Delta t] \}$ and he can evaluate $B_e$ as

$$B_e = B_o \sqrt{ 1 + \left[ {2 \over 3} {\Delta \Omega_{max} \over
{e B_o \over p} \Delta t } \right]^2 } \qquad . \eqno(A3)$$

\noindent
In this rough estimate we give the lower contribution from the irregular
magnetic field by assuming that trajectory perturbations are due to
structures of the wavelength $\lambda \sim c \Delta t$. In the case of
turbulence power concentrated at shorter waves the required wave power
is even
higher to cause the considered scattering. The turbulence power with
$\lambda > c \Delta t$ would provide correlations of successive
scatterings, excluded in the considered model.

{}
 
\end{document}